\documentclass[aps,preprint]{revtex4}%
\usepackage{amsfonts}
\usepackage{amsmath}
\usepackage{amssymb}
\usepackage{graphicx}%
\begin{document}
\preprint{ }
\title{The strategic form of quantum prisoners' dilemma}
\author{Ahmad Nawaz}
\affiliation{National Centre for Physics, Quaid-i-Azam University Campus}
\affiliation{Islamabad, Pakistan.}
\email{ahmad@ele.qau.edu.pk}

\begin{abstract}
In its normal form prisoners' dilemma (PD) is represented by a payoff matrix
showing players strategies and payoffs. To obtain distinguishing trait and
strategic form of PD certain constraints are imposed on the elements of its
payoff matrix. We quantize PD by generalized quantization scheme to analyze
its strategic behavior in quantum domain. The game starts with general
entangled state of the form $\left\vert \psi\right\rangle =\cos\frac{\xi}%
{2}\left\vert 00\right\rangle +i\sin\frac{\xi}{2}\left\vert 11\right\rangle $
and the measurement for payoffs is performed in entangled and product bases.
We show that for both measurements there exist respective cutoff values of
entanglement of initial quantum state up to which strategic form of game
remains intact. Beyond these cutoffs the quantized PD behaves like chicken
game up to another cutoff value. For the measurement in entangled basis the
dilemma is resolved for\ $\sin\xi>\frac{1}{7}$ with $Q\otimes Q$ as a NE but
the quantized game behaves like PD when $\sin\xi>\frac{1}{3}$; whereas in the
range $\frac{1}{7}<\sin\xi<\frac{1}{3}$ it behaves like chicken game
(CG)\ with $Q\otimes Q$ as a NE. For the measurement in product basis the
quantized PD behaves like classical PD for $\sin^{2}\frac{\xi}{2}<\frac{1}{3}$
with $D\otimes D$ as a NE. In region $\frac{1}{3}<\sin^{2}\frac{\xi}{2}%
<\frac{3}{7}$ the quantized PD behaves like classical CG with $C\otimes D$ and
$D\otimes C$ as NE.

\end{abstract}
\date{\today}
\maketitle

\section{Introduction}

Game theory deals with a situation where two or more rational players are
involved in a strategic contest to maximize their payoffs \cite{dixit}. The
payoff of each player depends on his own strategy and on the strategies
adopted by other players \cite{neumann}. The set of strategies from which
unilateral deviation of any player reduces his/ her payoff is called Nash
Equilibrium (NE) of the game \cite{nash}. In its normal form a game is
represented by a bi-matrix with its elements as payoffs. A set of constraints
is necessary to impose on the elements of the payoff matrix to obtain the
strategic form of the game. For example, prisoner dilemma (PD), which is a
story of two prisoners who have allegedly committed a crime together. They are
being interrogated in separate cells. Each of the\ prisoners have to decide
whether to confess the crime (to defect $D$) or to deny the crime (to
cooperate $C$) without any communication between them. If both players receive
$R$ and $U$ for mutual cooperation and defection respectively; and a
cooperator and defector engaged in a contest against each other receive $S$
and $T$ respectively; then the strategic form of PD demands that $T>R>U>S$
\cite{bengt,szabo}. Due these constraints rational reasoning forces each
player to defect. As a result $DD$ appears as a NE of the game which is not
\textit{Pareto optimal.} This is referred to as the dilemma of this game.

Chicken game (CG) on the other hand depicts a situation in which two players
drive their cars straight towards each other. The first to swerve to avoid the
collision $\left(  \text{to cooperate }C\right)  $ is the loser (chicken) and
the one who keeps on driving straight $\left(  \text{to defect }D\right)  $ is
the winner. By assigning $R$ and $U$ to mutual cooperation and defection
respectively; $S$ and $T$ to a cooperator and a defector against each other
then the strategic form of CG requires that $T>R>S>U$ \cite{bengt}. As a
result there is no dominant strategy and $CD$, $DC$ appear as NE. The dilemma
of this game is that $CC$ which is \textit{Pareto optimal} is not a NE.

This type of dilemmas was resolved by analyzing games in quantum domain. One
of the elegant and foremost step in this direction was by Eisert \textit{et al
}\cite{eisert} to remove dilemma in PD. In this quantization scheme the
strategy space of the players is a two parameter set of $2\times2$ unitary
operators. Starting with maximally entangled initial quantum state the authors
showed that for a suitable quantum strategy the dilemma disappears from the
game. The quantum strategy pair $Q\otimes Q$ appears as a NE which is
\textit{Pareto optimal}. They also pointed out that the quantum strategy $Q$
always wins over all classical strategies. Eisert \textit{et al}
\cite{eisert1} also showed that $Q\otimes Q$ is a unique NE in CG and is
\textit{Pareto optimal}. This quantization scheme has many interesting
applications in quantum game theory
\cite{du-2,zhou,lee,azhar,poit,flitney-1,cheon,ozdemir,ozdemir1,shimamura,chen,flitney-2,rosero,David}%
. Later on, Marinatto and Weber \cite{marinatto} introduced another
interesting and simple scheme for the quantization of non-zero sum games. They
gave Hilbert structure to the strategic spaces of the players. They also used
the maximally entangled initial state and allowed the players to play their
tactics by applying probabilistic choices of unitary operators. Applying their
scheme to Battle of Sexes game they found the strategy for which both the
players have equal payoffs. Marinatto and Weber quantization scheme gave very
interesting results while investigating evolutionarily stable strategies (ESS)
\cite{azhar,azhar1,ahmad02} and in the analysis of repeated games
\cite{azhar2,Piotr} etc.\emph{\ }In our earlier work we introduced a
generalized quantization scheme that establishes a relation between these two
apparently different quantization schemes \cite{ahmad}. Separate set of
parameters were identified for which this scheme reduces to that of Eisert
\textit{et al} \cite{eisert} and Marinatto and Weber \cite{marinatto}
quantization schemes.

In this paper we address the question that to what extent the strategic form
of PD remains unaffected if it is quantized by generalized quantization scheme
\cite{ahmad}. Starting with a general entangled state of the form $\left\vert
\psi\right\rangle =\cos\frac{\xi}{2}\left\vert 00\right\rangle +i\sin\frac
{\xi}{2}\left\vert 11\right\rangle $ we show that the strategic form of
quantized PD depends on entanglement of initial quantum state and as well as
on the type of measurement basis (entangled or product). For both types of
measurements there exist respective cutoff values of entanglement of initial
quantum state up to which strategic form of game remains intact. Beyond these
cutoffs the quantized PD behaves like chicken game up to another cutoff value.

The paper is organized as follows: section (\ref{pd cg}) is a brief
introduction to PD and CG, section (\ref{qpd}) presents that how the strategic
form of quantized PD\ changes by quantization and section (\ref{conc}%
)\ concludes the main results.

\section{\label{pd cg}Prisoners' Dilemma and Chicken Game}

Prisoner dilemma is the story of two suspects, Alice and Bob, who have
allegedly committed a crime together. They have been arrested and being
interrogated in separate cells. Each of the\ prisoners have to decide whether
to confess the crime or to deny the crime without any communication between
them. In game theory to confess the crime is termed as \textquotedblleft to
Defect\textquotedblright,the strategy $D$ and to deny the crime is referred to
as \textquotedblleft to Cooperate\textquotedblright, the strategy $C.$
Depending upon their decisions the players obtain the payoffs according the
the following payoff matrix.%

\begin{gather}%
\begin{array}
[c]{ccc}%
\text{ \ \ \ \ \ \ \ \ \ \ } & \text{Bob} &
\end{array}
\nonumber\\%
\begin{array}
[c]{cccc}%
\text{ \ \ \ \ \ \ \ \ \ \ } & \text{ \ \ \ }C & \text{ \ \ \ }D & \text{\ }%
\end{array}
\nonumber\\%
\begin{array}
[c]{c}%
\text{Alice}%
\end{array}%
\begin{array}
[c]{c}%
C\\
D
\end{array}
\left[
\begin{array}
[c]{cc}%
\left(  3,3\right)  & \left(  0,5\right) \\
\left(  5,0\right)  & \left(  1,1\right)
\end{array}
\right] \label{matrix}%
\end{gather}
It is clear from the above payoff matrix that $D$ is the dominant strategy for
both players. Therefore rational reasoning forces each of them to play $D $
resulting $DD$ as a NE of PD. From the payoff matrix (\ref{matrix}) we see
that each player gets $\left(  1,1\right)  $ as payoff. However, it was
possible for the players to get better payoff of of value $\left(  3,3\right)
$\ if they would have played $CC$ instead of $DD.$ This is generally known as
the dilemma of this game. We can write the payoff matrix (\ref{matrix}) in a
general form as
\begin{gather}%
\begin{array}
[c]{ccc}%
\text{ \ \ \ \ \ \ \ \ \ \ } & \text{Bob} &
\end{array}
\nonumber\\%
\begin{array}
[c]{cccc}%
\text{ \ \ \ \ \ \ \ \ \ \ } & \text{ \ \ \ }C & \text{ \ \ \ }D & \text{\ }%
\end{array}
\nonumber\\%
\begin{array}
[c]{c}%
\text{Alice}%
\end{array}%
\begin{array}
[c]{c}%
C\\
D
\end{array}
\left[
\begin{array}
[c]{cc}%
\left(  R,R\right)  & \left(  S,T\right) \\
\left(  T,S\right)  & \left(  U,U\right)
\end{array}
\right] \label{pd-general}%
\end{gather}
with
\begin{equation}
T>R>U>S\label{condition for pd}%
\end{equation}
as constraint on its elements.

In chicken game (CG) two players, Alice and Bob, drive their cars straight
towards each other. The first to swerve to avoid the collision $\left(
\text{the strategy }C\right)  $ is the loser (chicken) and the one who keeps
on driving straight $\left(  \text{the strategy }D\right)  $\ is the winner.
The payoff matrix for this game can also be of the form (\ref{pd-general}) but
with constraints
\begin{equation}
T>R>S>U.\label{condition for cg}%
\end{equation}
Certainly if both players cooperate they can avoid a crash and none of them
will be winner. If one of them steers away $\left(  \text{defects }D\right)  $
he will be loser but will survive but the opponent will receive the entire
honor. If they crash then the cost of both of them will be higher than the
cost of being chicken and the payoff will be lower \cite{bengt,russel}. There
is no dominant strategy in this game. The strategy pairs $\left(  C,D\right)
$ and $\left(  D,C\right)  $ are two NE in this game. The former is preferred
by Bob and the latter is preferred by Alice. The dilemma of this game is that
$CC$ which is \textit{Pareto optimal }is not the NE of this game.

\section{\label{qpd}Quantization of Prisoners' Dilemma}

In this section we quantize PD using generalized quantization scheme for two
person non zero sum games \cite{ahmad}. In this quantization scheme an arbiter
prepares a two qubit general entangled state and passes on one qubit to each
player. After applying their local unitary operators (strategies) the players
return the qubits to arbiter who then, announces the payoffs by performing the
measurement with the application of suitable payoff operators depending on the
payoff matrix of the game. The payoff operators are Bell like states which
transform to Eisert \textit{et al} \cite{eisert} operators for maximum
entanglement and for zero entanglement they reduce to the payoff operators
used by Marinatto and Weber in their quantization scheme \cite{marinatto}.
There can be four cases of interest. If both the initial quantum state and
payoff operators are in form of product states then classical game is
reproduced. When initial quantum state and the payoff operators are maximally
entangled states then this scheme transforms to Eisert \textit{et al
}\cite{eisert}\textit{\ }quantization scheme. For maximally entangled initial
quantum state and product basis measurement it is reduced to that of Marinatto
and Weber quantization scheme \cite{marinatto}. On the other hand if the game
starts with product state but the measurement for the payoffs is performed in
entangled basis then the payoffs are also quantum mechanical in nature. Where
as this feature is absent both in Eisert \textit{et al} and Marinatto and
Weber quantization schemes.

For the quantization of PD the classical strategies $C$ (to cooperate) and $D
$ (to defect) are assigned two basis vectors $\left\vert C\right\rangle $ and
$\left\vert D\right\rangle $ respectively in a Hilbert space of two level
system. The state of game at any instant is a vector in four dimensional
Hilbert space spanned by the basis vectors $\left\vert CC\right\rangle ,$
$\left\vert CD\right\rangle ,$ $\left\vert DC\right\rangle $ and $\left\vert
DD\right\rangle .$ Here the entries in the ket refer to the qubits possessed
by Alice and Bob respectively. Representing $\left\vert C\right\rangle
\rightarrow\left\vert 0\right\rangle $ and $\left\vert D\right\rangle
\rightarrow\left\vert 1\right\rangle $ let the initial quantum state of game
be of the form
\begin{equation}
\left\vert \psi\right\rangle =\cos\frac{\xi}{2}\left\vert 00\right\rangle
+i\sin\frac{\xi}{2}\left\vert 11\right\rangle \label{input state}%
\end{equation}
where $\xi$ is the entanglement parameter. The strategies of players are
represented by unitary operators $U_{j},$ given as \cite{ahmad}%
\begin{equation}
U_{j}=\cos\frac{\theta_{j}}{2}R_{j}+\sin\frac{\theta_{j}}{2}C_{j}%
\label{strategy operator}%
\end{equation}
where $j=A,B$ and $R_{j},$ $C_{j}$ are the unitary operators defined as \ %

\begin{align}
R_{j}\left\vert 0\right\rangle  & =e^{i\phi_{j}}\left\vert 0\right\rangle
,\text{ \ }R_{j}\left\vert 1\right\rangle =e^{-i\phi_{j}}\left\vert
1\right\rangle \nonumber\\
C_{j}\left\vert 0\right\rangle  & =-\left\vert 1\right\rangle ,\text{ \ }%
C_{j}\left\vert 1\right\rangle =\left\vert 0\right\rangle .\label{strategies}%
\end{align}
After the application of strategies the initial state given by Eq.
(\ref{input state}) transforms into
\begin{equation}
\rho_{f}=\left(  U_{A}\otimes U_{B}\right)  \rho\left(  U_{A}\otimes
U_{B}\right)  ^{\dagger}.\label{state after strategies}%
\end{equation}
The payoff operators of Alice and Bob are
\begin{align}
P^{A}  & =3P_{00}+P_{11}+5P_{10}\nonumber\\
P^{B}  & =3P_{00}+P_{11}+5P_{01}\label{payoff operators}%
\end{align}
where
\begin{subequations}
\label{projector}%
\begin{align}
P_{00}  & =\left\vert \psi_{00}\right\rangle \left\langle \psi_{00}\right\vert
,\text{ \ }\left\vert \psi_{00}\right\rangle =\cos\frac{\delta}{2}\left\vert
00\right\rangle +i\sin\frac{\delta}{2}\left\vert 11\right\rangle \label{a}\\
P_{11}  & =\left\vert \psi_{11}\right\rangle \left\langle \psi_{11}\right\vert
,\text{ \ }\left\vert \psi_{11}\right\rangle =\cos\frac{\delta}{2}\left\vert
11\right\rangle +i\sin\frac{\delta}{2}\left\vert 00\right\rangle \label{b}\\
P_{10}  & =\left\vert \psi_{10}\right\rangle \left\langle \psi_{10}\right\vert
,\text{ \ }\left\vert \psi_{10}\right\rangle =\cos\frac{\delta}{2}\left\vert
10\right\rangle -i\sin\frac{\delta}{2}\left\vert 01\right\rangle \label{c}\\
P_{01}  & =\left\vert \psi_{01}\right\rangle \left\langle \psi_{01}\right\vert
,\text{ \ }\left\vert \psi_{01}\right\rangle =\cos\frac{\delta}{2}\left\vert
01\right\rangle -i\sin\frac{\delta}{2}\left\vert 10\right\rangle \label{d}%
\end{align}
and $\delta\in\left[  0,\frac{\pi}{2}\right]  $ is the entanglement of
measurement basis. These payoff operators reduce to that of Eisert \textit{et
al} scheme \cite{eisert} for $\delta=\frac{\pi}{2}$ and for $\delta=0$ these
transform to that of Marinatto and Weber scheme \cite{marinatto}. The payoff
for player $i$ are calculated as
\end{subequations}
\begin{equation}
\$_{i}\left(  \theta_{A},\phi_{A},\theta_{B},\phi_{B}\right)  =\text{Tr}%
\left(  P^{i}\rho_{f}\right)  .\label{trace}%
\end{equation}
Since in generalized quantization scheme measurements can be performed in
entangled as well as in product basis therefore we discuss both the cases one
by one.

\textbf{Case 1:- Entangled measurement}

\begin{figure}
[h]
\begin{center}
\includegraphics[
height=2.1715in,
width=3.3797in
]%
{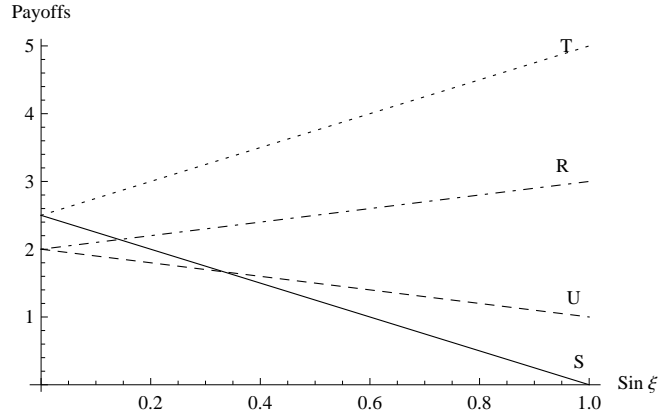}%
\caption{Payoff elements versus $\sin\xi$ when the measurement is performed in
entngled basis. It shows that the quantized PD behaves like PD for an initial
quantum state for which $\sin\xi>\frac{1}{3}$. In the region $\frac{1}{3}%
\geq\sin\xi\geq\frac{1}{7}$ the quantized PD behaves like CG and for $\sin
\xi<\frac{1}{7}$ the form of the game is again changed.}%
\label{entangled01}%
\end{center}
\end{figure}

When the measurement is performed in entangled basis then using Eqs.
(\ref{matrix}, \ref{input state}, \ref{state after strategies},
\ref{payoff operators}, \ref{trace}) the payoffs of players come out to be%

\begin{align}
\$_{A}\left(  \theta_{A},\phi_{A},\theta_{B},\phi_{B}\right)   & =\left[
2+\sin\xi\cos2\left(  \phi_{A}+\phi_{B}\right)  \right]  \cos^{2}\frac
{\theta_{A}}{2}\cos^{2}\frac{\theta_{B}}{2}\nonumber\\
& +\frac{5}{2}\left(  1+\sin\xi\cos2\phi_{B}\right)  \sin^{2}\frac{\theta_{A}%
}{2}\cos^{2}\frac{\theta_{B}}{2}\nonumber\\
& +\frac{5}{2}\left(  1-\sin\xi\cos2\phi_{A}\right)  \cos^{2}\frac{\theta_{A}%
}{2}\sin^{2}\frac{\theta_{B}}{2}\nonumber\\
& +\left(  2-\sin\xi\right)  \sin^{2}\frac{\theta_{A}}{2}\sin^{2}\frac
{\theta_{B}}{2}\nonumber\\
& -\frac{\left(  2+\sin\xi\right)  }{4}\sin\theta_{A}\sin\theta_{B}\sin\left(
\phi_{A}+\phi_{B}\right) \nonumber\\
& -\frac{5}{4}\sin\theta_{A}\sin\theta_{B}\sin\left(  \phi_{A}-\phi
_{B}\right)  .\label{payoff a}%
\end{align}%
\begin{align}
\$_{B}\left(  \theta_{A},\phi_{A},\theta_{B},\phi_{B}\right)   & =\left[
2+\sin\xi\cos2\left(  \phi_{A}+\phi_{B}\right)  \right]  \cos^{2}\frac
{\theta_{A}}{2}\cos^{2}\frac{\theta_{B}}{2}\nonumber\\
& +\frac{5}{2}\left(  1+\sin\xi\cos2\phi_{A}\right)  \sin^{2}\frac{\theta_{B}%
}{2}\cos^{2}\frac{\theta_{A}}{2}\nonumber\\
& +\frac{5}{2}\left(  1-\sin\xi\cos2\phi_{B}\right)  \cos^{2}\frac{\theta_{B}%
}{2}\sin^{2}\frac{\theta_{A}}{2}\nonumber\\
& +\left(  2-\sin\xi\right)  \sin^{2}\frac{\theta_{A}}{2}\sin^{2}\frac
{\theta_{B}}{2}\nonumber\\
& -\frac{\left(  2+\sin\xi\right)  }{4}\sin\theta_{A}\sin\theta_{B}\sin\left(
\phi_{A}+\phi_{B}\right) \nonumber\\
& -\frac{5}{4}\sin\theta_{A}\sin\theta_{B}\sin\left(  \phi_{A}-\phi
_{B}\right)  .\label{payoff b}%
\end{align}
In this case if the game starts from maximally entangled state then $Q\otimes
Q$ is the only NE of the game\ where $Q$ is the unitary operator $U\left(
\theta,\phi\right)  =U\left(  0,\frac{\pi}{2}\right)  $ \cite{eisert}. To see
the behavior of $Q\otimes Q$ at other values of entanglement we apply the NE
conditions as
\begin{align}
\$_{A}\left(  0,\frac{\pi}{2},0,\frac{\pi}{2}\right)  -\$_{A}\left(
\theta_{A},\phi_{A},0,\frac{\pi}{2}\right)   & \geq0\nonumber\\
\$_{B}\left(  0,\frac{\pi}{2},0,\frac{\pi}{2}\right)  -\$_{B}\left(
0,\frac{\pi}{2},\theta_{B},\phi_{B}\right)   & \geq0.\label{NE condition}%
\end{align}
With the help of Eqs. (\ref{payoff a}, \ref{payoff b}) for $i=A,B$ the above
inequalities give%

\begin{equation}
7\sin\xi+\left[  1+\left(  2\cos2\phi_{i}-5\right)  \sin\xi\right]  \cos
^{2}\frac{\theta_{i}}{2}-1\geq0.\text{ }\label{inequality gamma}%
\end{equation}
This inequality is satisfied for\ $\sin\xi\geq\frac{1}{7}$. Therefore
$Q\otimes Q$ remains NE for a game that starts with an initial state for
which
\begin{equation}
\sin\xi>\frac{1}{7}.\label{value of xi}%
\end{equation}
Now we investigate whether the quantum game that we obtained by quantization
of PD with $Q\otimes Q$ as NE possesses the characteristics of PD. Using Eqs.
(\ref{payoff a}, \ref{payoff b}) the elements of payoff matrix of quantized PD
are
\begin{equation}
R=2+\sin\xi,\text{ }S=\frac{5-5\sin\xi}{2},\text{ }T=\frac{5+5\sin\xi}%
{2},\text{ }U=2-\sin\xi.\text{ }\label{rstu-entangled}%
\end{equation}
For the above values of payoff elements the constraint (\ref{condition for pd}%
) is satisfied if $\sin\xi>\frac{1}{3}$. It shows that in quantized PD the
resolution of dilemma without effecting its strategic form requires that the
entanglement of initial quantum state must be greater than $\arcsin\frac{1}%
{3}.$ It means that $Q\otimes Q$\ is the NE of a quantized PD for all values
of entanglement for which $\sin\xi\geq\frac{1}{7}$\ but it behaves like PD
only for $\sin\xi>\frac{1}{3}.$ This is shown in figure (\ref{entangled01}).
It is evident from figure (\ref{entangled01}) that in the region $\frac{1}%
{3}\geq\sin\xi\geq\frac{1}{7}$ the constraints on payoff elements transforms
to
\begin{equation}
T>R>S>U.
\end{equation}
Comparing with (\ref{condition for cg}) we see that for these values of
entanglement the quantized PD behaves like CG. It means that when PD is
quantized with an initial state of entanglement less than $\arcsin\frac{1}{3}$
then it transforms to CG but $Q\otimes Q$\ still remains the NE.\ When the
entanglement is further reduced and $\sin\xi<\frac{1}{7}$ then the quantum
game again changes its form. In this region the payoff matrix elements obey
the constraints
\begin{equation}
T>S>R>U
\end{equation}
and $Q\otimes D$, $D\otimes Q$ are NE. For $Q\otimes D$ the NE conditions
\begin{align}
\$_{A}\left(  Q,D\right)  -\$_{A}\left(  \theta_{A},\phi_{A},D\right)   &
\geq0\nonumber\\
\$_{B}\left(  Q,D\right)  -\$_{B}\left(  Q,\theta_{B},\phi_{B}\right)   &
\geq0
\end{align}
become%
\begin{align}
\sin^{2}\frac{\theta_{A}}{2}+\left[  7-\left(  2-5\cos2\phi_{A}\right)
\cos^{2}\frac{\theta_{A}}{2}\right]  \sin\xi & \geq0\nonumber\\
\left[  1-\left(  5-2\cos2\phi_{B}\right)  \sin\xi\right]  \cos^{2}%
\frac{\theta_{B}}{2}  & \geq0
\end{align}
These inequalities are satisfied for all $\theta^{\prime}s$ and $\phi^{\prime
}s$ if $0\leq\sin\xi\leq\frac{1}{7}.$ However at $\sin\xi=0$ two new NE
$C\otimes D$ and $D\otimes C$ also come into play. At this stage we have a
game that has four pure strategies NE. For $C\otimes D$ the NE conditions
\begin{align}
\$_{A}\left(  C,D\right)  -\$_{A}\left(  \theta_{A},\phi_{A},D\right)   &
\geq0\nonumber\\
\$_{B}\left(  C,D\right)  -\$_{B}\left(  C,\theta_{B},\phi_{B}\right)   &
\geq0
\end{align}
yield%
\begin{align}
\sin^{2}\frac{\theta_{A}}{2}-\left[  3+\left(  2-5\cos2\phi_{A}\right)
\cos^{2}\frac{\theta_{A}}{2}\right]  \sin\xi & \geq0\nonumber\\
\left[  1+\left(  5-2\cos2\phi_{B}\right)  \sin\xi\right]  \cos^{2}%
\frac{\theta_{B}}{2}  & \geq0.
\end{align}
These inequalities are satisfied for all $\theta^{\prime}s$ and $\phi^{\prime
}s$ for $\sin\xi=0$ showing that for zero value of entanglement $C\otimes D$
becomes a NE.

On the other hand it is also obvious from figure (\ref{entangled01}) there are
two points $\sin\xi=\frac{1}{7}$ and $\sin\xi=\frac{1}{3}$ where the
constraints on the elements of the payoff matrix are
\begin{equation}
T>R=S>U
\end{equation}
and
\begin{equation}
T>R>S=U
\end{equation}
respectively. The former constraint represents the game called compromise
dilemma \cite{bengt} whereas the other constraint represents a game that is
also different than PD.

%

It proves that when PD starts with a general entangled state of the form
(\ref{input state}) and measurement is performed in entangled basis then it
behaves like PD up to a certain cutoff value of entanglement of initial
quantum state.

\textbf{Case 2:- Product Measurement}

When the measurement is performed in product basis then using Eqs.
(\ref{matrix}, \ref{input state}, \ref{state after strategies},
\ref{payoff operators}, \ref{trace}) the payoff of player A comes out to be%

\begin{align}
\$_{A}\left(  \theta_{A},\phi_{A},\theta_{B},\phi_{B}\right)   & =\left(
1+2\cos^{2}\frac{\xi}{2}\right)  \cos^{2}\frac{\theta_{A}}{2}\cos^{2}%
\frac{\theta_{B}}{2}+5\cos^{2}\frac{\xi}{2}\sin^{2}\frac{\theta_{A}}{2}%
\cos^{2}\frac{\theta_{B}}{2}\nonumber\\
& +5\sin^{2}\frac{\xi}{2}\cos^{2}\frac{\theta_{A}}{2}\sin^{2}\frac{\theta_{B}%
}{2}+\left(  1+2\sin^{2}\frac{\xi}{2}\right)  \sin^{2}\frac{\theta_{A}}{2}%
\sin^{2}\frac{\theta_{B}}{2}\nonumber\\
& -\frac{1}{4}\sin\xi\sin\theta_{A}\sin\theta_{B}\sin\left(  \phi_{A}+\phi
_{B}\right)  .\label{payoffa product}%
\end{align}
The payoffs of player B can be found by replacing $\theta_{A}\rightarrow
\theta_{B}$ and $\phi_{A}\rightarrow\phi_{B}.$ For these payoffs $C\otimes C$
is the NE of the game if
\begin{equation}
\$_{A}\left(  C,C\right)  -\$_{A}\left(  \theta_{A},\phi_{A},C\right)  \geq0.
\end{equation}
Putting the corresponding values from Eq. (\ref{payoffa product}) we get
\begin{equation}
\sin^{2}\frac{\theta_{A}}{2}\left(  3\sin^{2}\frac{\xi}{2}-2\right)
\geq0.\label{NE product}%
\end{equation}
This inequality is satisfied if $\sin^{2}\frac{\xi}{2}\geq\frac{2}{3}.$ It
shows that $C\otimes C$ is a NE with payoff $3-2\sin^{2}\frac{\xi}{2}$\ for
quantized PD that starts with an initial entangled state of the form
(\ref{input state})\ if $\sin^{2}\frac{\xi}{2}\geq\frac{2}{3}$. It is
important to note that $C\otimes C$ being a NE does not imply the resolution
of dilemma. Because for $\sin^{2}\frac{\xi}{2}\geq\frac{2}{3}$ each player
could have obtained a better payoff $1+2\sin^{2}\frac{\xi}{2}$ by playing $D$
instead of $C.$

For $D\otimes D$ as a NE we have the inequality
\begin{equation}
\$_{A}\left(  D,D\right)  -\$_{A}\left(  \theta_{A},\phi_{A},D\right)  \geq0,
\end{equation}
which with help of Eq. (\ref{payoffa product}) gives
\begin{equation}
\cos^{2}\frac{\theta_{A}}{2}\left(  1-3\sin^{2}\frac{\xi}{2}\right)  \geq0.
\end{equation}
The above inequality is satisfied for $\sin^{2}\frac{\xi}{2}\leq\frac{1}{3}$
showing that $D\otimes D$ is a NE for quantized PD if it starts with an
initial quantum state with $\sin^{2}\frac{\xi}{2}\leq\frac{1}{3}.$

The $D\otimes C$ can be NE if it satisfies the following NE inequalities
\begin{align}
\$_{A}\left(  D,C\right)  -\$_{A}\left(  \theta_{A},\phi_{A},C\right)   &
\geq0\nonumber\\
\$_{B}\left(  D,C\right)  -\$_{B}\left(  D,\theta_{B},\phi_{B}\right)   &
\geq0.
\end{align}
With the help of Eq. (\ref{payoffa product}) the above inequalities become
\begin{align}
\cos^{2}\frac{\theta_{A}}{2}\left(  2-3\sin^{2}\frac{\xi}{2}\right)   &
\geq0\nonumber\\
\sin^{2}\frac{\theta_{B}}{2}\left(  3\sin^{2}\frac{\xi}{2}-1\right)   & \geq0.
\end{align}
These inequalities are satisfied if $\frac{1}{3}\leq\sin^{2}\frac{\xi}{2}%
\leq\frac{2}{3}.$ Therefore, $D\otimes C$ is NE of quantized PD which starts
with an initial entangled state with $\frac{1}{3}\leq\sin^{2}\frac{\xi}{2}%
\leq\frac{2}{3}.$ By similar reasoning it can be proved that $C\otimes D$ is
NE if the entanglement of initial quantum state is in the range $\frac{1}%
{3}\leq\sin^{2}\frac{\xi}{2}\leq\frac{2}{3}.$

\begin{figure}
[ptbh]
\begin{center}
\includegraphics[
height=2.1292in,
width=3.3797in
]%
{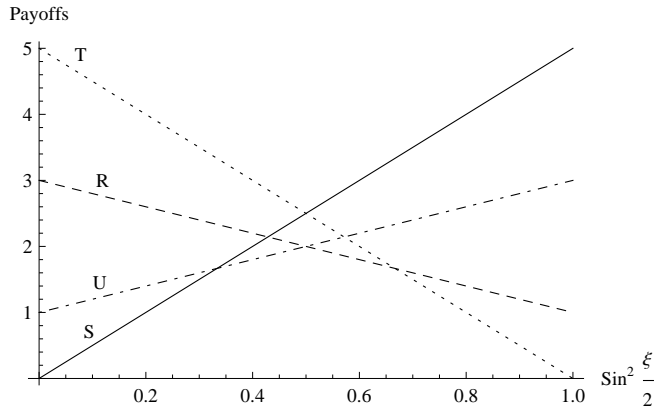}%
\caption{Payoff elements versus $\sin^{2}\frac{\xi}{2}$ when the measurement
is performed in product basis. The constraints required for the game to behave
like PD are satisfied in region defined by $0\leq\sin^{2}\frac{\xi}{2}%
<\frac{1}{3}$. In the region $\frac{1}{3}<\sin^{2}\frac{\xi}{2}<\frac{3}{7}$
the quantized PD behaves classical CG with $C\otimes D$ and $D\otimes C$ as
NE.}%
\label{product01}%
\end{center}
\end{figure}

Now we investigate that how the strategic form of quantized PD depends upon
the entanglement of initial state. We find from Eq. (\ref{payoffa product})
that the elements of payoff matrix in this case are
\begin{equation}
R=3-2\sin^{2}\frac{\xi}{2},S=5\sin^{2}\frac{\xi}{2},T=5-5\sin^{2}\frac{\xi}%
{2},U=1+2\sin^{2}\frac{\xi}{2}.\text{ }\label{elements product}%
\end{equation}
These elements of the payoff matrix are plotted as a function of $\sin
^{2}\frac{\xi}{2}$ in figure (\ref{product01}). The figure shows six regions
and each region represents a different game. The constraints
(\ref{condition for pd}) required for the game to behave like PD are satisfied
in region 1. This region is defined as $0\leq\sin^{2}\frac{\xi}{2}<\frac{1}%
{3}$ with $D\otimes D$ as the NE. In region 2 where $\frac{1}{3}<\sin^{2}%
\frac{\xi}{2}<\frac{3}{7}$ the payoff matrix elements given in Eq.
(\ref{elements product}) are transformed into the constraints given in
(\ref{condition for cg}). In this region the quantized PD represents classical
CG with $C\otimes D$ and $D\otimes C$ as NE. When the entanglement of initial
state is further increased then form of the game varies according to table
(\ref{table}).%

\begin{table}[h] \centering
\[%
\begin{tabular}
[c]{|l|l|l|l|}\hline
Region & Entanglement & NE & Game\\\hline
1 & $\sin^{2}\frac{\xi}{2}<\frac{1}{3}$ & $D\otimes D$ & Classical PD\\\hline
2 & $\frac{1}{3}<\sin^{2}\frac{\xi}{2}<\frac{3}{7}$ & $C\otimes D$, $D\otimes
C$ & Classical CG\\\hline
3 & $\frac{3}{7}<\sin^{2}\frac{\xi}{2}<\frac{1}{2}$ & $C\otimes D$, $D\otimes
C$ & Neither CG nor PD\\\hline
4 & $\frac{1}{2}<\sin^{2}\frac{\xi}{2}<\frac{4}{7}$ & $C\otimes D$, $D\otimes
C$ & Neither CG nor PD\\\hline
5 & $\frac{4}{7}<\sin^{2}\frac{\xi}{2}<\frac{2}{3}$ & $C\otimes D$, $D\otimes
C$ & Neither CG nor PD\\\hline
6 & $\sin^{2}\frac{\xi}{2}>\frac{2}{3}$ & $C\otimes C$ & Neither CG nor
PD\\\hline
\end{tabular}
\]
\caption{ Different forms of PD for specified range of initial state
entanglement when measurement is performed in product basis}\label{table}%
\end{table}%

Furthermore from figure (\ref{product01}) it can be seen that there are points
such as $\sin^{2}\frac{\xi}{2}=\frac{1}{3},$ $\sin^{2}\frac{\xi}{2}=\frac
{3}{7},$ $\sin^{2}\frac{\xi}{2}=\frac{1}{2},$ $\sin^{2}\frac{\xi}{2}=\frac
{4}{7}$ and $\sin^{2}\frac{\xi}{2}=\frac{2}{3}$ where two or more payoff
matrix elements are equal. The form of game at these points can be described
as follows.

\begin{enumerate}
\item At $\sin^{2}\frac{\xi}{2}=\frac{1}{3}$ the payoff matrix elements are
related through the constraints $T>R>S=U$ with\ $R=2.3333,$ $S=U=1.6667,$ and
$T=3.3333.$ Here the game has $C\otimes D$ and $D\otimes C$ as NE. But both
these NE are not strict \cite{martin}$.$

\item At $\sin^{2}\frac{\xi}{2}=\frac{3}{7}$ we see that $T>\left(
R=S\right)  >U$ with $R=S=2.1429,$ $T=2.8571,$ $U=1.8571$ and at this point
quantized PD behaves like compromise dilemma. In such a situation it is better
to play opposite to the opponent \cite{bengt}.

\item At $\sin^{2}\frac{\xi}{2}=\frac{1}{2}$ the matrix elements obey the
constraints $\left(  R=U\right)  <\left(  S=T\right)  $ where $R=U=2,$
$S=T=2.5.$ Here the game has $C\otimes D$ and $D\otimes C$ as NE.

\item At $\sin^{2}\frac{\xi}{2}=\frac{4}{7}$ the constraints on the payoff
matrix element become $S>\left(  T=U\right)  >R$ and $R=1.8571,$ $S=2.8571,$
$T=U=2.1429.$ The game has $C\otimes D$ and $D\otimes C$ as NE.

\item At $\sin^{2}\frac{\xi}{2}=\frac{2}{3}$ the constraints take the form
$S>U>\left(  R=T\right)  $ where $R=T=1.6667,$ $S=3.3333,$ $U=2.3333.$ This
game has $C\otimes D$\textrm{\ }and\textrm{\ }$D\otimes C$\textrm{\ }as
NE\textrm{\ }and both these NE are not strict \cite{martin}.
\end{enumerate}

Note that for all the above cases the quantized game never behaves like PD.%

\section{\label{conc}Conclusion}

We quantized PD by generalized quantization scheme \cite{ahmad} starting with
a general initial entangled state of the form $\left\vert \psi\right\rangle
=\cos\frac{\xi}{2}\left\vert 00\right\rangle +i\sin\frac{\xi}{2}\left\vert
11\right\rangle .$ In this scheme the measurements for payoffs can be
performed in entangled and product bases. For both types of measurements the
strategic form of quantized PD depends upon the entanglement of initial
quantum state. For measurement in entangled basis when the entanglement of
initial quantum state is reduced then beyond a certain level of entanglement
the quantized PD behaves like CG with $Q\otimes Q$ as NE. On further reduction
of entanglement the game ceased to behave like CG and transformed into a new
game with $Q\otimes D$ and $D\otimes Q$ as NE. At last for zero entanglement
two additional NE $C\otimes D$ and $D\otimes C$ also appeared resulting a game
with four pure strategies NE. When the measurement is performed in product
basis then the quantized PD can be divided in eleven different games with
respect to initial state entanglement. In this case for zero entanglement of
initial quantum state the game behaved like PD with $D\otimes D$ as NE. With
increasing entanglement of initial state there is a cutoff value beyond that
game behaved like CG with $C\otimes D$ and $D\otimes C$ as NE. On further
increase there appeared another cutoff value beyond that the quantized PD
transformed into a game with $C\otimes C$ as NE which is not \textit{Pareto
optimal}.

The apparent reason for these results is when the players apply their pure
strategies ( $I$ and $\sigma$ operators) on a maximally entangled state (Bell
state) shared between them then the resulting quantum state is also one of the
Bell states. This state overlaps with one of the payoff operators
(\ref{projector}) and is orthogonal to other three operators. Therefore the
measurement of payoffs is is error free. However, when the entanglement of
shared quantum state is reduced then application of pure strategies transform
it into a state which overlaps with two payoff operators (\ref{projector}).
The payoffs against the pure strategies ( $I$ and $\sigma$ operators) are
transformed into the payoffs corresponding to mixed strategies (linear
combination of $I$ and $\sigma$ ). It changes the strategic form of the game.
Similarly the case of product measurements can be explained.

\end{document}